\documentclass[
reprint,
superscriptaddress,
aps,
amsmath,amssymb,
pra,
]{revtex4-1}

\usepackage{graphicx}
\usepackage{color}
\usepackage{setspace}
\usepackage{array}
\usepackage[version=3]{mhchem}
\usepackage{nth}

\begin{document}

\title{Electron Localization and Energy Levels' Oscillations Induced by Controlled Deformation}

\author{Fahhad H. Alharbi}%
\email{falharbi@qf.org.qa}
\affiliation{Qatar Environment and Energy Research Institute (QEERI), Qatar Foundation, Doha, Qatar}

\author{Pablo Serra}%
\affiliation{Qatar Environment and Energy Research Institute (QEERI), Qatar Foundation, Doha, Qatar}
\affiliation{Facultad de Matem\'atica, Astronom\'{\i}a y F\'{\i}sica, Universidad Nacional de C\'ordoba and IFEG-CONICET, Ciudad Universitaria, X5016LAE C\'ordoba, Argentina}

\author{Marcelo A Carignano}%
\affiliation{Qatar Environment and Energy Research Institute (QEERI), Qatar Foundation, Doha, Qatar}

\author{Sabre Kais}%
\affiliation{Qatar Environment and Energy Research Institute (QEERI), Qatar Foundation, Doha, Qatar}
\affiliation{Department of Chemistry, Physics, and Birck Nanotechnology Center, Purdue University, West Lafayette, Indiana 47907, USA)}

\begin{abstract}
Manipulating energy levels while controlling the electron localization is an essential step for many applications of confined systems. In this paper we demonstrate how to achieve electron localization and induce energy level oscillation in one-dimensional quantum systems by externally controlling the deformation of the system. From a practical point of view, the one-dimensional potentials can be realized  using layered structures. In the analysis, we considered three different examples. The first one is a graded quantum well between confining infinite walls where the deformation is modeled by varying slightly the graded well. The second systems is a symmetric multiple quantum well between infinite walls under the effect of biasing voltage. The third system is a layered 2D hybrid perovskites where pressure is used to induce deformation. The calculations are conducted both numerically and analytically using the perturbation theory. It is shown that the obtained oscillations are associated with level avoided crossings and that the deformation results in changing the spatial localization of the electrons. 
\end{abstract}

\maketitle

\section{Introduction}

Over the past decade, manipulating electrons in quantum-confined system has progressed remarkably \cite{Y01,Y02} and thus allows developing devices for many applications like quantum computing \cite{S06,V02}, spectroscopy \citep{A02,B02}, ultra-fast and sensitive optoelectronics \cite{N02,Z04}, optical tunability \cite{E01,L03}, and negative heat capacity \cite{S10}. In such devices, it is essential to have well controlled manipulation means. Different manipulation mechanisms are used currently such as superconductor circuits \cite{O01,S01} and spins in solids \cite{H01} and molecules \cite{V01,C01,O02}. However, the controllability of confined quantum systems in general and in particular for selected energy levels is still a major challenge \cite{S02,F01} that has stimulating the exploration of many new concepts \cite{A01,S03,Y03,Z01,Z02}. One of the largely investigated control techniques relies on the utilization of St\"{u}ckelberg oscillations. By passing an avoided crossing twice, the interference due to the dynamical phase between transitions can be either constructive or destructive. This is known as Landau-Zener-St\"{u}ckelberg (LZS) interference \cite{S01,S03,S04}. Usually, the levels' oscillation occurs in the time domain. However, Nori and co-workers showed that this is achievable as well in the space domain using spatially inhomogeneous magnetic field \cite{Z03}. Furthermore, similar oscillatory behaviors originated from different external influences  and not intended for LZS interferometry, have been reported in other types of systems. For example, using self assembled quantum dots \cite{S07} under a strain field, quantum dots in micro cavities under terahertz laser excitation \cite{S08} or coupled semiconductor nano rings with varying inter-ring distance \cite{C03}.

In this paper we present practical methods to manipulate electron localization and to cause energy levels' oscillation in quantum confined systems using controlled deformations that can, in principle, be applied at room temperature. These methods can be used to realized LZS interferometry where the oscillation is caused by various means. By deforming the quantum system, the system potential energy is slightly altered and hence the states and their energies are changed accordingly. The practical application of these ideas depends on the magnitude of the level oscillation and displacement of the localization centers, and how these are coupled to the externally controlled deformations. We found that the control of the state localization is easily achievable with deformations. The control of the energy levels requires careful engineering addressing the binding strength of the electron to particular sites that can be obtained using, for example, quantum dots. The assumed controlled deformations are applied either by voltage biasing or pressure across 2D layered systems such as graded semiconductor quantum wells \cite{L04,H03,B04} or stacks of 2D hybrid perovskites \cite{E01,M02}. Thus, the problem becomes a piece-wise constant potential and hence it has an analytical solution \cite{F02,S05} where the eigenfunctions, which are exponential and trigonometric functions with continuous logarithmic derivative, and the eigenenergies are obtained as solutions of transcendental algebraic equations. Also, the problem can be solved numerically using various methods like finite difference (FDM) \cite{H02,F03}, finite element (FEM) \cite{N01,L01}, and spectral methods \cite{F04,L02}. In this work, we use a flexible FDM \cite{F03} to calculate the eigen-pairs. 

The numerical calculations show that energy level oscillations are achievable and manipulatable in the studied systems by controlled deformations. Furthermore, the obtained oscillations are associated with avoided crossings. Thus, it can be used to realize LZS interferometry using alternative means other than the time-varying field. It is shown also that beside the oscillations, the deformation results in changing the spatial localization of the electrons. This should provide means to manipulate the electrons and hence can be used for other applications of quantum-confined systems. The next section presents three different scenarios of controlled deformations, followed by a final section with our general conclusions.

\section{Controlled deformations}

\begin{figure} [t]
\begin{center}
\includegraphics[width=3.2in]{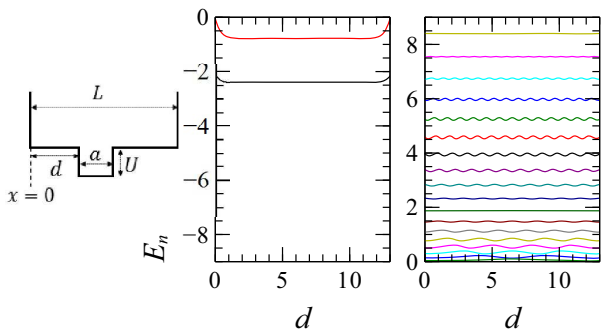}
\end{center}
\caption{  \label{Fig01} (top left) The studied single-well structure. (bottom left and right panels) Numerical  calculation of the  first 20 eigenenergies for the example system ($L=15,\,a=2,\,U=3$). The eigenenergies less than zero are shown in the bottom left panel while those greater than zero are shown in the right panel.}
\end{figure}

In order to explore possible scenarios in which localization/delocalization transitions occur via controlled deformations and how are they linked to energy level oscillations we studied three different confined systems. The first one is a graded quantum well between infinite walls where the deformation is modeled by varying spatially the graded potential. The second studied systems is a symmetric multiple quantum well between infinite walls under the effect of biasing voltage. The third system is a layered 2D hybrid perovskites where the pressure is used to induce deformation. Tuning the properties through pressure has been applied for purely inorganic systems \cite{B04,M01}. However, the softer character of organic layers should lead to more pronounced effects in 2D hybrid materials \cite{E01,L03}. For the calculations, we use a flexible finite FDM \cite{F03} for the calculations as aforementioned and atomic units are assumed throughout the paper.

\subsection{Graded quantum well between infinite walls}

The system analyzed in this subsection is a square well between two infinite walls separated by a distance $L$. So, the potential (Figure-\ref{Fig01}) is simply
\begin{equation}
\label{pot}
V(x)\,=\,\left\{ \begin{array}{lll} 
0 & \mbox{if } & 0 < x < d \; \mbox{or} \; d+a < x < L \\
-U & \mbox{if } & d < x < d+a \\
\infty &\mbox{otherwise} \\
 \end{array}  \right. \,,
\end{equation}
where $L,\,d,\, a$ and $U$ are all positive constants and $d+a < L$. The potential deformation is modeled by varying $d$, which is the distance between the left infinite wall and the edge of the well. Namely, the deformation represents to displace the position of the potential well with respect to the confining walls. As an example of the effects we show some of the results obtained with $L=15$, $a=2$, and $U=3$. In this case, we obtain two states with energy levels below zero (shown in lower left of Figure-\ref{Fig01}) that are bounded stated to the small well. The energies above zero as a function of $d$ are shown in the right panel of Figure-\ref{Fig01}, up to the 20th level. It is clear that these states are oscillating with $d$ and that the amplitude of the oscillations depends not in a monotonic way with the order of the level.

\begin{figure*} [t]
\begin{center}
\includegraphics[width=4.5in]{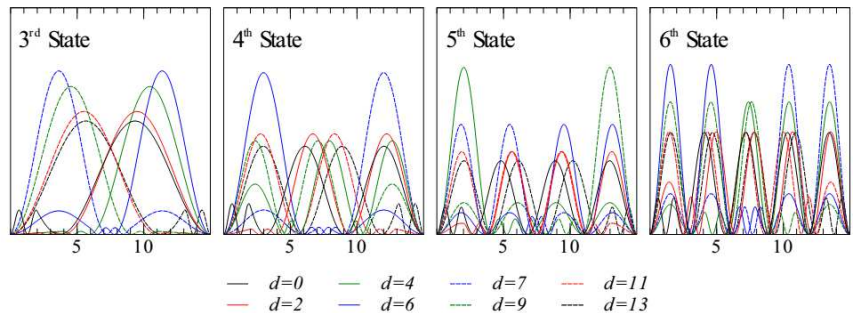}
\end{center}
\caption{  \label{Fig06} The densities for different states as indicated and selected values if $d$. As the small well is displaced from left to right there is a concurrent displacement of the density that can be concentrated on either side depending on the state level being observed.}
\end{figure*}

In order to have an analytical perspective that could shed light into the different contributions that result on the observed oscillatory behavior, the non degenerate eigenenergies (above zero in the used model) are calculated using perturbation theory (PT). In this case, the unperturbed structure is assumed to be an infinite walls potential and the perturbation is due to the small finite well. So, the Hamiltonian can be rewritten as $H=T+V_0+V_1$, where 
\begin{equation}
\label{hppV0}
V_0(r)\,=\,\left\{ \begin{array}{lll}
0 & \mbox{if } & 0 < x < L \\
\infty &\mbox{otherwise} & \\
 \end{array}  \right.
\end{equation}
and
\begin{equation}
\label{hppV1}
V_1(r)\,=\,\left\{ \begin{array}{lll}
-U & \mbox{if}& \;d< x <d+ a \\
0 & \mbox{otherwise } &\mbox{}\\
 \end{array}  \right. 
\end{equation}
The well-known eigenvalues of $H_0$ ($T+V_0$) are  $E^{(0)}_m= \frac{1}{2} k_m^2 = \frac{1}{2}\left( \frac{m \pi}{L} \right)^2$, where 
$m=1, 2, \ldots$, is the sum of peaks and valleys of the corresponding eigenfunctions, which are 
\begin{equation}
\label{efo0}
\psi_m(x) \, = \, \sqrt{\dfrac{2}{L}} \sin{(k_m x)} \,\,\,\,\,\,\, .
\end{equation}
Here, $m$ is used in this subsection as the counter for the states with energies above zero and should not be confused with $n$ that counts all the states. With the exception of this subsection, $n$ is used to describe the energy levels throughout this paper. The first order correction to the energies is given by
\begin{equation}
\begin{split}
\label{foce}
\Delta^{(1)}_m\,&= \langle m \left| V_1 \right| m \rangle = -U\,\int_d^{d+a}\,\psi_m^2(x) \,dx \\
&= -U \dfrac{a}{L} + U \dfrac{1}{m \pi} \sin{\left( m \pi \dfrac{a}{L} \right)} \cos{\left( m \pi \dfrac{(2d+a)}{L} \right)} \,\,.
\end{split}
\end{equation}
It is clear from the above equation that the corrections are oscillating with $d$ at a frequency of $2 m \pi/L$, which matches the oscillations obtained by the numerical calculations especially for higher levels. Eq.-\ref{foce} also implies that the oscillation amplitude is proportional to $U$ and is sinusoidal with $m$ and $a$, providing additional means to design and control the oscillations. In particular, the amplitude dependency on $m$ explains the behavior observed on Figure \ref{Fig01} showing stronger oscillations at intermediate values of $m$.

\begin{figure} [b]
\begin{center}
\includegraphics[width=3.4in]{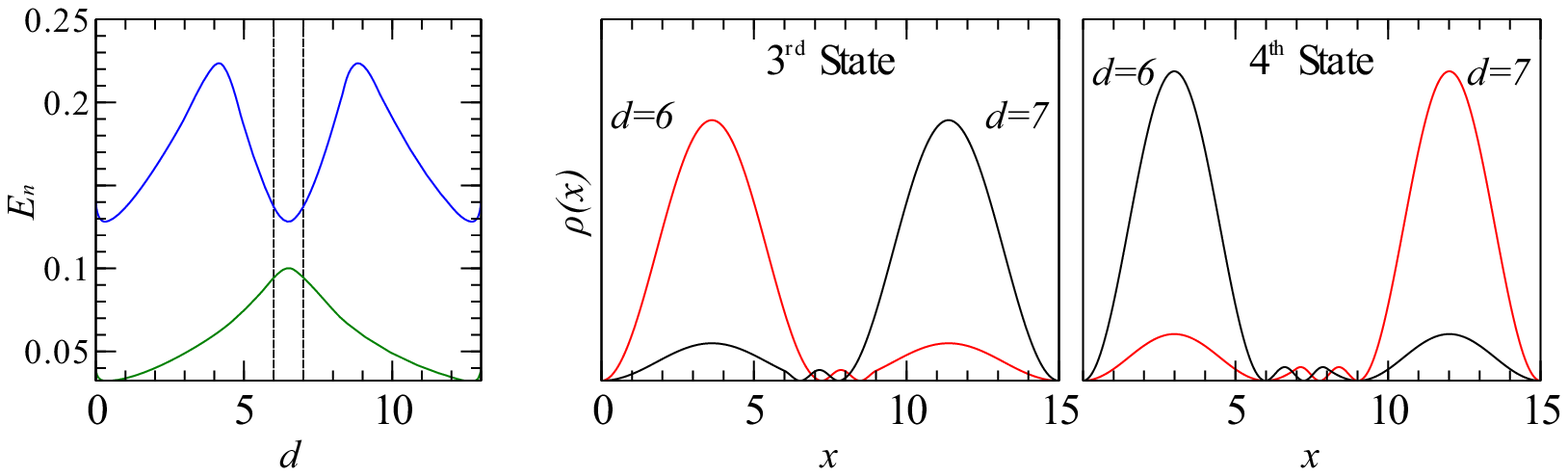}
\end{center}
\caption{  \label{Fig07} (color online) (left) The energies for the third and forth eignestates vs. $d$. (middle) The density for the third eigenstate for $d=6$ and $d=7$. (right) The density for the firth eigenstate for $d=6$ and $d=7$.}
\end{figure}

The localization of the electron, as described by the density $\rho(x)=|\psi(x)|^2$, is also affected by the deformation.  In Figure \ref{Fig06} we show the resulted densities for the first four positive eigenenergies, $n=$3, 4, 5, and 6, for a selection of different $d$ values. There is a clear shifting from one side of the system to the other on the most probable position of the particle as a function of the system state. For example, for the third state, the particle is localized in the right side of the well for $d < (L-a)/2$. The localization is shifted toward the left side for $d > (L-a)/2$. This shifting effect persists up to higher states, although the side chosen by the system depends, among other factor, on the odd/even character of the quantum number.

So far we have shown that energy levels' oscillations (for energies above zero) are directly related to the system's deformation and therefore providing a way to externally control the particular values of the energy spectrum of a system. As aforementioned in the introduction section, many similar oscillations, due to other causes, were reported and in many of them the oscillations are associated with crossing avoidance. The oscillations observed in this example are also associated with density swapping between states. In Figure \ref{Fig07}, the energies and densities of the third and forth eigenstates are shown. In the left panel, it is clear that there is a crossing avoidance at $d=6.5$. This is associated with a density swapping where for $d=6$, the third state is localized in the left of the well while the fourth state is localized in the right. This swapping happens in the reverse order for $d=7$.

\subsection{A symmetric multiple quantum well between infinite walls deformed by biasing}

\begin{figure} [t]
\begin{center}
\includegraphics[width=3.3in]{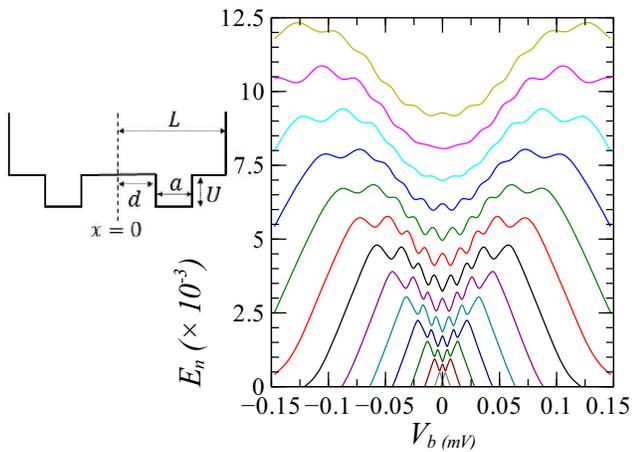}
\end{center}
\caption{  \label{Fig08} The sketch shows the double-well symmetric structure. The left figure displays the energy levels as a function of the biasing voltage $V_b$ and the induced oscillations. The different colors represent the different energy levels.}
\end{figure}

In this subsection and the next we consider two examples that could be practically achieved with currently known materials and technologies, allowing the experimental exploration of controlled energy level oscillations and the associated localization/delocalization of the particles. In this case, we start with a symmetric double-well system confined between two infinite walls. Then, a biasing voltage is applied to deform the original system into a tilted one. The unbiased potential, represented in the sketch of Figure \ref{Fig08}, is mathematically described by the following potential:
\begin{equation}
\label{pot}
V(x)\,=\,\left\{ \begin{array}{lll} 
0 & \mbox{if } & |x|< d \;\mbox{or} \;d+a\leq |x| < L \\
-U & \mbox{if}& \;d<|x| <d+ a \\
\infty
& \mbox{otherwise} \\
 \end{array}  \right. \,,
\end{equation}
where $L,\,d,\, a$ and $U$ are positive constants and $d+a < L$. Equivalent systems can be made although with finite potential walls. For example, using semiconductor multiple quantum wells \cite{L05,C04} or a double quantum dot  system \cite{S11,H04} can be used to realize a device having the qualitative features of Eq. (\ref{pot}). Having this examples in mind, the parameters used in this analysis are within the ranges of practical values. We chose $L=200$, $a=40$ and $U=0.02$ that represent ~10 nm, ~2 nm and ~0.54 eV, respectively. By applying a biasing voltage $V_b$, the potential is tilted by an additional term of $-V_b \, x$ providing the externally controlled deformation. In our example, the voltage is varied between -0.15 and 0.15 mV. The resulting energies are shown in Figure \ref{Fig08} as a function of $V_b$, where it can be seen the emergence of an oscillating pattern having similar characteristics to the example of the previous section. For example, the amplitude of the oscillation is a function of the energy level, as it was found by the PT analysis presented above.

\begin{figure} [t]
\begin{center}
\includegraphics[width=3.4in]{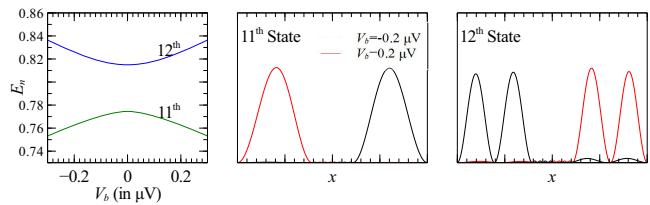}
\end{center}
\caption{  \label{Fig09b} (left) The energies for the 11\textsuperscript{th} and 12\textsuperscript{th} eignestates vs. $V_b$. (middle) The density for the 11\textsuperscript{th} eigenstate for $V_b=-0.2$ $\mu$V and $V_b=0.2$ $\mu$V. (right) The density for the 12\textsuperscript{th} eigenstate for $V_b=-0.2$ $\mu$V and $V_b=0.2$ $\mu$V.}
\end{figure}

Analogously to the case of Subsection A, in this example the energy oscillation has an associated crossing avoidance and localization transition. In Figure \ref{Fig09b} we show the energies and densities corresponding to the 11\textsuperscript{th} and 12\textsuperscript{th} eigenstates. In the left panel, it is clear that there is a crossing avoidance around $V_b=0$ mV. The concurrent density swapping happens upon a very small change in the biasing: for $V_b=-0.01$ mV, the 11\textsuperscript{th} state is localized in the left of the well while the 12\textsuperscript{th} state is localized in the right. The picture is completely reversed for $V_b=0.01$ mV.

\subsection{A layered 2D hybrid perovskites deformed by applying pressure}

The second possible practical deformation method that we consider relies on the application of an external pressure. To realize this system we propose a layered 2D hybrid perovskites. These structures are composed of alternating layers of hybrid 2D framework of octahedra inorganic and organic cations providing the wells. Moreover, the system could be intercalated between softer polymeric layers allowing large deformations and increasing the degree of external control by the application of moderate pressure as compared with purely inorganic crystals \cite{F05,A03,S09}. These materials can be analyzed as superlattices, as shown by Even et al. \cite{E01} and the references therein.

\begin{figure}[t]
\begin{center}
\includegraphics[origin=c,width=2.75in]{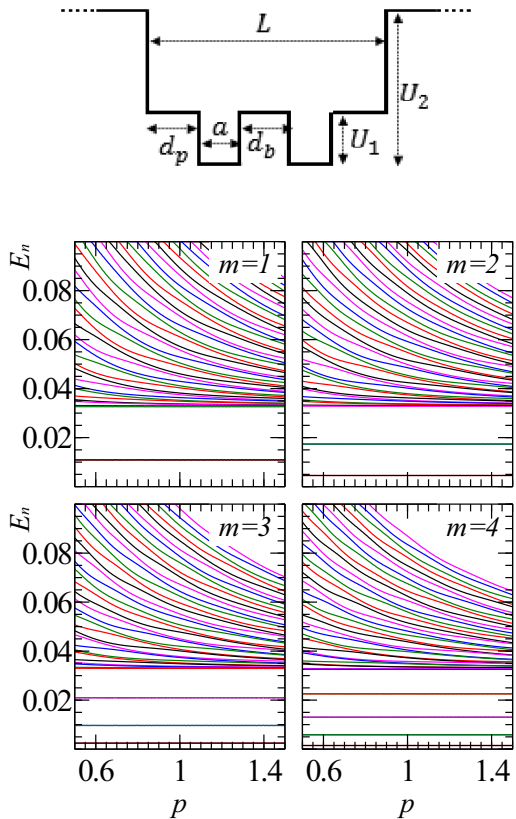}
\end{center}
\caption{  \label{Fig10} (top) Model potential to represent 2D layer perovskites confined between insulating walls, in this case for two layers creating two confining wells. The deformation affects the external confining layers of width $d_p$. (bottom) Numerical calculation of the first 40 eigenenergies for a two wells system with $m=$1, 2, 3, and 4 as a function of the deformation factor.}
\end{figure}

The most studied 2D hybrid materials are \ce{[RNH3]2(CH3NH3)_{m-1}Pb_{m}I_{3m+1}}, with \ce{R} representing an organic group. In these materials, \ce{Pb_{m}I_{3m+1}} forms the octahedra that include the \ce{(CH3NH3)} cations forming the wells. On the other hand, \ce{[RNH3]2} constitutes the barriers and by changing the organic group (R), the equilibrium separation between the wells can be tailored \cite{E01}. The depth of the well is estimated to be $U_1=0.9$ eV \cite{T01}. As for the well width, it depends on the number of the octahedra layers. For $m=$1, 2, 3, and 4, the thicknesses $a$ are 0.65, 1.25, 1.85, and 2.5 nm respectively \cite{T01}. 

\begin{figure} [t]
\begin{center}
\includegraphics[width=3.5in]{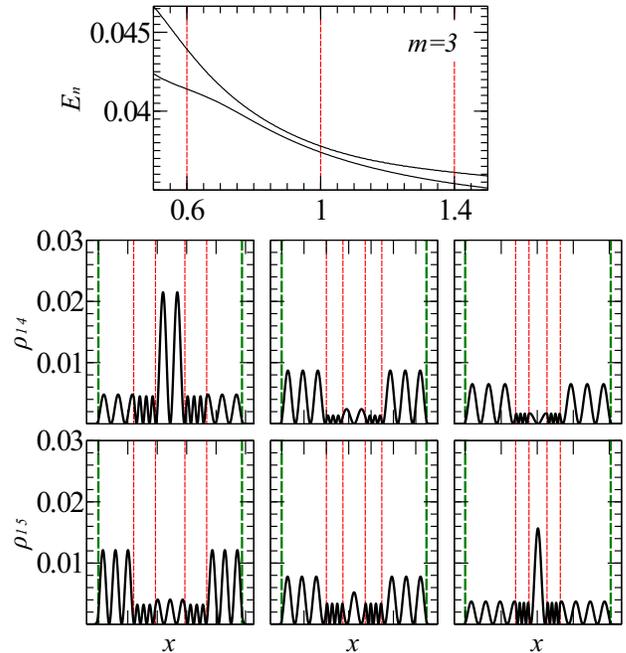}
\end{center}
\caption{  \label{Fig11} (top) Eigenvalues corresponding the 14\textsuperscript{th} and 15\textsuperscript{th} energy levels for the case of two wells and $m=3$ as function of the deformation factor (the pressure $p$). The red vertical dashed lines correspond to the the selected deformations for which the densities are plotted for the two states as function of distance. (bottom) The densities of the two states as function of distance where the dashed green lines are corresponding to the edges of the insulating walls while the dashed red lines are corresponding to the small wells edges.}
\end{figure}

In this subsection, we study the effect of the pressure on the energy levels on the \ce{[RNH3]2(CH3NH3)_{m-1}Pb_{m}I_{3m+1}} quantum wells with 1.25 nm barrier. A finite number of wells will be used where the whole system is then placed between two insulators with a barrier height of $U_2=2.7$ eV, as described in Figure \ref{Fig10} for the case of two wells \cite{E01}. In our approximation, the external pressure only affects the confining external layers by changing their thickness by a factor $p$. The eigenvalues show the avoidance crossing characteristics as in the previous cases, although now as a function of the deformation parameter $p$. This is explicitly displayed in Figure \ref{Fig11}, in which we display two states corpsonding to the case of $m=3$. As the system is deformed, the 14\textsuperscript{th} energy level has the particle shifting its preferential position from the central region between the wells for $p=0.6$, to the outermost layers for $p=1.4$. The energy level immediately above, the situation is exactly the opposite.

\section{Conclusion}

In this paper, we have analyzed the interplay between the detailed dimensions of a confining potential, energy level oscillations and particle's localization. It was shown that the obtained oscillations are associated with level avoided crossings and that the deformation induces a change in the preferential spatial localization of the electrons. This effect is quite general, as shown by the the three described examples, and therefore it can be realized in different ways. The external controlling parameter could be an applied voltage or pressure, and other procedures are certainly possible. Obtaining a practical way to experimentally control the electron localization would enable the manipulation of charges in quantum confined systems and provide guidelines for the design of devices.
The approach that we have used to describe the is based on simple 1D problems, but the consistent presence of this effect on multiple scenarios suggests that this phenomena is general and could be realized in 3D systems at the nanometer scale.

\section{References}

\bibliographystyle{unsrt}
\bibliography{EnOscP01}

\begin{thebibliography}{10}

\bibitem{Y01}
JR~Petta, AC~Johnson, JM~Taylor, EA~Laird, A~Yacoby, MD~Lukin, CM~Marcus,
  MP~Hanson, and AC~Gossard.
\newblock Coherent manipulation of coupled electron spins in semiconductor
  quantum dots.
\newblock {\em Science}, 309(5744):2180--2184, 2005.

\bibitem{Y02}
JM~Taylor, JR~Petta, AC~Johnson, A~Yacoby, CM~Marcus, and MD~Lukin.
\newblock Relaxation, dephasing, and quantum control of electron spins in
  double quantum dots.
\newblock {\em Physical Review B}, 76(3):035315, 2007.

\bibitem{S06}
RJ~Schoelkopf and SM~Girvin.
\newblock Wiring up quantum systems.
\newblock {\em Nature}, 451(7179):664--669, 2008.

\bibitem{V02}
D~Vion, A~Aassime, A~Cottet, Pl~Joyez, H~Pothier, C~Urbina, D~Esteve, and
  Michel~H Devoret.
\newblock Manipulating the quantum state of an electrical circuit.
\newblock {\em Science}, 296(5569):886--889, 2002.

\bibitem{A02}
O~Astafiev, Alexandre~M Zagoskin, AA~Abdumalikov, Yu~A Pashkin, T~Yamamoto,
  K~Inomata, Y~Nakamura, and JS~Tsai.
\newblock Resonance fluorescence of a single artificial atom.
\newblock {\em Science}, 327(5967):840--843, 2010.

\bibitem{B02}
David~M Berns, Mark~S Rudner, Sergio~O Valenzuela, Karl~K Berggren, William~D
  Oliver, Leonid~S Levitov, and Terry~P Orlando.
\newblock Amplitude spectroscopy of a solid-state artificial atom.
\newblock {\em Nature}, 455(7209):51--57, 2008.

\bibitem{N02}
Thomas Niemczyk, F~Deppe, H~Huebl, EP~Menzel, F~Hocke, MJ~Schwarz,
  JJ~Garcia-Ripoll, D~Zueco, T~H{\"u}mmer, E~Solano, et~al.
\newblock Circuit quantum electrodynamics in the ultrastrong-coupling regime.
\newblock {\em Nature Physics}, 6(10):772--776, 2010.

\bibitem{Z04}
M~Zecherle, C~Ruppert, EC~Clark, G~Abstreiter, JJ~Finley, and M~Betz.
\newblock Ultrafast few-fermion optoelectronics in a single self-assembled in
  ga as/gaas quantum dot.
\newblock {\em Physical Review B}, 82(12):125314, 2010.

\bibitem{E01}
Jacky Even, Laurent Pedesseau, and Claudine Katan.
\newblock Understanding quantum confinement of charge carriers in layered 2d
  hybrid perovskites.
\newblock {\em ChemPhysChem}, 15(17):3733--3741, 2014.

\bibitem{L03}
Ga{\"e}tan Lanty, Khaoula Jemli, Yi~Wei, Jo{\"e}l Leymarie, Jacky Even,
  Jean-S{\'e}bastien Lauret, and Emmanuelle Deleporte.
\newblock Room-temperature optical tunability and inhomogeneous broadening in
  2d-layered organic--inorganic perovskite pseudobinary alloys.
\newblock {\em The Journal of Physical Chemistry Letters}, 5(22):3958--3963,
  2014.

\bibitem{S10}
Pablo Serra, Marcelo~A. Carignano, Fahhad~H. Alharbi, and Sabre Kais.
\newblock Quantum confinement and negative heat capacity.
\newblock {\em Europhysics Letters}, 104(1):16004, 2013.

\bibitem{O01}
William~D Oliver, Yang Yu, Janice~C Lee, Karl~K Berggren, Leonid~S Levitov, and
  Terry~P Orlando.
\newblock Mach-zehnder interferometry in a strongly driven superconducting
  qubit.
\newblock {\em Science}, 310(5754):1653--1657, 2005.

\bibitem{S01}
Mika Sillanp{\"a}{\"a}, Teijo Lehtinen, Antti Paila, Yuriy Makhlin, and Pertti
  Hakonen.
\newblock Continuous-time monitoring of landau-zener interference in a
  cooper-pair box.
\newblock {\em Physical review letters}, 96(18):187002, 2006.

\bibitem{H01}
Ronald Hanson and David~D Awschalom.
\newblock Coherent manipulation of single spins in semiconductors.
\newblock {\em Nature}, 453(7198):1043--1049, 2008.

\bibitem{V01}
Lieven~MK Vandersypen and Isaac~L Chuang.
\newblock Nmr techniques for quantum control and computation.
\newblock {\em Reviews of modern physics}, 76(4):1037, 2005.

\bibitem{C01}
Ben Criger, Gina Passante, Daniel Park, and Raymond Laflamme.
\newblock Recent advances in nuclear magnetic resonance quantum information
  processing.
\newblock {\em Philosophical Transactions of the Royal Society A: Mathematical,
  Physical and Engineering Sciences}, 370(1976):4620--4635, 2012.

\bibitem{O02}
Sangchul Oh, Zhen Huang, Uri Peskin, and Sabre Kais.
\newblock Entanglement, berry phases, and level crossings for the atomic
  breit-rabi hamiltonian.
\newblock {\em Physical Review A}, 78(6):062106, 2008.

\bibitem{S02}
Sonia~G Schirmer, Ivan~CH Pullen, and Allan~I Solomon.
\newblock Controllability of multi-partite quantum systems and selective
  excitation of quantum dots.
\newblock {\em Journal of Optics B: Quantum and Semiclassical Optics},
  7(10):S293, 2005.

\bibitem{F01}
Alejandro Ferr{\'o}n, Pablo Serra, and Omar Osenda.
\newblock Quantum control of a model qubit based on a multi-layered quantum
  dot.
\newblock {\em Journal of Applied Physics}, 113:134304, 2013.

\bibitem{A01}
David~D Awschalom, Lee~C Bassett, Andrew~S Dzurak, Evelyn~L Hu, and Jason~R
  Petta.
\newblock Quantum spintronics: Engineering and manipulating atom-like spins in
  semiconductors.
\newblock {\em Science}, 339(6124):1174--1179, 2013.

\bibitem{S03}
SN~Shevchenko, Sahel Ashhab, and Franco Nori.
\newblock Landau--zener--st{\"u}ckelberg interferometry.
\newblock {\em Physics Reports}, 492(1):1--30, 2010.

\bibitem{Y03}
Norman~Y Yao, Liang Jiang, Alexey~V Gorshkov, Peter~C Maurer, Geza Giedke,
  J~Ignacio Cirac, and Mikhail~D Lukin.
\newblock Scalable architecture for a room temperature solid-state quantum
  information processor.
\newblock {\em Nature Communications}, 3:800, 2012.

\bibitem{Z01}
Jingfu Zhang, Man-Hong Yung, Raymond Laflamme, Al{\'a}n Aspuru-Guzik, and
  Jonathan Baugh.
\newblock Digital quantum simulation of the statistical mechanics of a
  frustrated magnet.
\newblock {\em Nature Communications}, 3:880, 2012.

\bibitem{Z02}
Yu-Jing Zhao, Xi-Ming Fang, Fang Zhou, and Ke-Hui Song.
\newblock Scheme for realizing quantum-information storage and retrieval from
  quantum memory based on nitrogen-vacancy centers.
\newblock {\em Physical Review A}, 86(5):052325, 2012.

\bibitem{S04}
SN~Shevchenko, S~Ashhab, and Franco Nori.
\newblock Inverse landau-zener-st{\"u}ckelberg problem for qubit-resonator
  systems.
\newblock {\em Physical Review B}, 85(9):094502, 2012.

\bibitem{Z03}
J-N Zhang, C-P Sun, S~Yi, and Franco Nori.
\newblock Spatial landau-zener-st{\"u}ckelberg interference in spinor
  bose-einstein condensates.
\newblock {\em Physical Review A}, 83(3):033614, 2011.

\bibitem{S07}
Weidong Sheng and Jean-Pierre Leburton.
\newblock Anomalous quantum-confined stark effects in stacked inas/gaas
  self-assembled quantum dots.
\newblock {\em Physical review letters}, 88(16):167401, 2002.

\bibitem{S08}
Mark~S Sherwin, Atac Imamoglu, and Thomas Montroy.
\newblock Quantum computation with quantum dots and terahertz cavity quantum
  electrodynamics.
\newblock {\em Physical Review A}, 60(5):3508, 1999.

\bibitem{C03}
T~Chwiej and B~Szafran.
\newblock Few-electron artificial molecules formed by laterally coupled quantum
  rings.
\newblock {\em Physical Review B}, 78(24):245306, 2008.

\bibitem{L04}
LC~Lenchyshyn, HC~Liu, M~Buchanan, and ZR~Wasilewski.
\newblock Voltage-tuning in multi-color quantum well infrared photodetector
  stacks.
\newblock {\em Journal of applied physics}, 79(10):8091--8097, 1996.

\bibitem{H03}
Alexander H{\"o}gele, Stefan Seidl, Martin Kroner, Khaled Karrai, Richard~J
  Warburton, Brian~D Gerardot, and Pierre~M Petroff.
\newblock Voltage-controlled optics of a quantum dot.
\newblock {\em Physical review letters}, 93(21):217401, 2004.

\bibitem{B04}
W~Bardyszewski and SP~{\L}epkowski.
\newblock Pressure-dependent reordering of valence band states in gan/al x ga
  1- x n quantum wells.
\newblock {\em Physical Review B}, 85(3):035318, 2012.

\bibitem{M02}
David~B Mitzi, Konstantinos Chondroudis, and Cherie~R Kagan.
\newblock Organic-inorganic electronics.
\newblock {\em IBM journal of research and development}, 45(1):29--45, 2001.

\bibitem{F02}
Siegfried Fl{\"u}gge.
\newblock {\em Practical quantum mechanics}.
\newblock Springer Verlag, 1994.

\bibitem{S05}
Pablo Serra and Sabre Kais.
\newblock Ground-state stability and criticality of two-electron atoms with
  screened coulomb potentials using the b-splines basis set.
\newblock {\em Journal of Physics B: Atomic, Molecular and Optical Physics},
  45(23):235003, 2012.

\bibitem{H02}
Paul Harrison.
\newblock {\em {Quantum Wells, Wires and Dots: Theoretical and Computational
  Physics}}.
\newblock John Wiley \& Sons, Chichester, 1st edition, 2000.

\bibitem{F03}
Fahhad Alharbi.
\newblock An explicit fdm calculation of nonparabolicity effects in energy
  states of quantum wells.
\newblock {\em Optical and quantum electronics}, 40(8):551--559, 2008.

\bibitem{N01}
Kenji Nakamura, Akira Shimizu, Masanori Koshiba, and Kazuya Hayata.
\newblock Finite-element analysis of quantum wells of arbitrary semiconductors
  with arbitrary potential profiles.
\newblock {\em Quantum Electronics, IEEE Journal of}, 25(5):889--895, 1989.

\bibitem{L01}
Khai~Q Le.
\newblock Finite element analysis of quantum states in layered quantum
  semiconductor structures with band nonparabolicity effect.
\newblock {\em Microwave and Optical Technology Letters}, 51(1):1--5, 2009.

\bibitem{F04}
Fahhad Alharbi.
\newblock Meshfree eigenstate calculation of arbitrary quantum well structures.
\newblock {\em Physics Letters A}, 374(25):2501--2505, 2010.

\bibitem{L02}
Wenbin Lin, Narayan Kovvali, and Lawrence Carin.
\newblock Pseudospectral method based on prolate spheroidal wave functions for
  semiconductor nanodevice simulation.
\newblock {\em Computer physics communications}, 175(2):78--85, 2006.

\bibitem{M01}
R~Meyer, M~Dahl, G~Schaack, A~Waag, and R~Boehler.
\newblock Low-temperature magneto-optical studies of a cdte/cd 1- x mn x te
  quantum-well structure at high hydrostatic pressures.
\newblock {\em Solid state communications}, 96(5):271--278, 1995.

\bibitem{L05}
Matthew~P Lumb, Michael~K Yakes, Mar{\'\i}a Gonz{\'a}lez, Igor Vurgaftman,
  Christopher~G Bailey, Raymond Hoheisel, and Robert~J Walters.
\newblock Double quantum-well tunnel junctions with high peak tunnel currents
  and low absorption for inp multi-junction solar cells.
\newblock {\em Applied Physics Letters}, 100(21):213907, 2012.

\bibitem{C04}
Gabriel Christmann, Alexis Askitopoulos, George Deligeorgis, Zacharias
  Hatzopoulos, Simeon~I Tsintzos, Pavlos~G Savvidis, and Jeremy~J Baumberg.
\newblock Oriented polaritons in strongly-coupled asymmetric double quantum
  well microcavities.
\newblock {\em Applied Physics Letters}, 98(8):081111, 2011.

\bibitem{S11}
Charles~A Stafford and Ned~S Wingreen.
\newblock Resonant photon-assisted tunneling through a double quantum dot: An
  electron pump from spatial rabi oscillations.
\newblock {\em Physical review letters}, 76(11):1916, 1996.

\bibitem{H04}
Xuedong Hu and S~Das Sarma.
\newblock Hilbert-space structure of a solid-state quantum computer:
  Two-electron states of a double-quantum-dot artificial molecule.
\newblock {\em Physical Review A}, 61(6):062301, 2000.

\bibitem{F05}
Jarvist~M Frost, Keith~T Butler, Federico Brivio, Christopher~H Hendon, Mark
  van Schilfgaarde, and Aron Walsh.
\newblock Atomistic origins of high-performance in hybrid halide perovskite
  solar cells.
\newblock {\em Nano Letters}, 14(5):2584--2590, 2014.

\bibitem{A03}
RJ~Angel, J~Zhao, and NL~Ross.
\newblock General rules for predicting phase transitions in perovskites due to
  octahedral tilting.
\newblock {\em Physical review letters}, 95(2):025503, 2005.

\bibitem{S09}
Raymond~E Schaak and Thomas~E Mallouk.
\newblock Perovskites by design: a toolbox of solid-state reactions.
\newblock {\em Chemistry of Materials}, 14(4):1455--1471, 2002.

\bibitem{T01}
Kenichiro Tanaka and Takashi Kondo.
\newblock Bandgap and exciton binding energies in lead-iodide-based natural
  quantum-well crystals.
\newblock {\em Science and Technology of Advanced Materials}, 4(6):599--604,
  2003.

\end{thebibliography}

\end{document}